\documentclass[aip, rip, reprint]{revtex4-1}
\usepackage{graphicx}

\begin{document}


\title{High-finesse nanofiber Fabry-P\'{e}rot resonator in a portable storage container} 



\author{S. Horikawa}
\affiliation{Nanofiber Quantum Technologies, Inc., 1-22-3 Nishiwaseda, Shinjuku-ku, Tokyo 169-0051, Japan.}
\affiliation{Department of Applied Physics, Waseda University, 3-4-1 Okubo, Shinjuku-ku, Tokyo 169-8555, Japan.}

\author{S. Yang}
\affiliation{Nanofiber Quantum Technologies, Inc., 1-22-3 Nishiwaseda, Shinjuku-ku, Tokyo 169-0051, Japan.}
\affiliation{Department of Applied Physics, Waseda University, 3-4-1 Okubo, Shinjuku-ku, Tokyo 169-8555, Japan.}

\author{T. Tanaka}
\affiliation{Nanofiber Quantum Technologies, Inc., 1-22-3 Nishiwaseda, Shinjuku-ku, Tokyo 169-0051, Japan.}

\author{T. Aoki}
\affiliation{Nanofiber Quantum Technologies, Inc., 1-22-3 Nishiwaseda, Shinjuku-ku, Tokyo 169-0051, Japan.}
\affiliation{Department of Applied Physics, Waseda University, 3-4-1 Okubo, Shinjuku-ku, Tokyo 169-8555, Japan.}

\author{S. Kato}
\affiliation{Nanofiber Quantum Technologies, Inc., 1-22-3 Nishiwaseda, Shinjuku-ku, Tokyo 169-0051, Japan.}
\email[]{shinya.kato@nano-qt.com}



\date{\today}

\begin{abstract}
We present characterization and storage methods for a high-finesse nanofiber Fabry-P\'{e}rot resonator.
Reflection spectroscopy from both ends of the resonator allows for evaluation of the mirror transmittances and optical loss inside the resonator.
To maintain the quality of the nanofiber resonator after the fabrication, we have developed a portable storage container.
By filling the container with dry, clean nitrogen gas, we can prevent contamination of the nanofiber during storage.
This approach allows us to minimize the additional optical loss to less than 0.08\% over a week.
The portable container facilitates both the fabrication and subsequent experimentation with the resonator in different locations. This flexibility expands the range of applications, including quantum optics, communication, and sensing.
\end{abstract}


\maketitle 

\section{Introduction}
The optical silica nanofiber, which has a diameter smaller than the wavelength of propagating laser light, serves as an optical waveguide that facilitates low-loss optical transmission and acts as an interface between the propagating mode and external entities such as atoms or molecules via the evanescent field\cite{Brambilla2010-nj, Morrissey2013-ra}. Its straightforward fabrication process\cite{Ward2014-bq}, in contrast to conventional photonic chip manufacturing techniques, and its seamless integration with standard fiber optics through fiber taper have led to a rapid expansion of its applications, spanning from physical and optical sciences\cite{Ismaeel2013-od,Tong2018-fh} to biomedical fields\cite{Praveen_Kamath2023-wl}.

For quantum science applications, optical fiber resonators, including nanofibers, enhance the interaction between the resonator mode and the interacting atoms, thereby constituting a cavity quantum electrodynamics (CQED) system\cite{Kato2015-ax, Ruddell2017-uo}. To enhance the device's ability to collect, guide, and control the resonator photon, it is important to reduce loss inside the resonator, especially at the nanofiber and fiber taper region. From a practical usability perspective, maintaining the quality of the nanofiber is crucial, as degradation prior to device installation in a vacuum or cryogenic environment can affect its functionality\cite{Fujiwara2011-hv, Bouhadida2021-gb}.

In this work, we have developed characterization and storage methods for a high-finesse optical nanofiber resonator.
The resonator is fabricated from a standard single-mode optical fiber by inscribing a fiber Bragg grating (FBG) mirror pair and creating a nanofiber region using a heat-and-pull method.
Optimizing the fabrication process including the heating condition results in an observed finesse of 3.6(2)$\times 10^3$, which is about two-fold improvement over previously reported values\cite{Ruddell2020-vb}.
The loss inside the resonator is determined with reflection spectroscopy from both ends of the resonator, yielding a round-trip loss of 0.14(1)\%.

In order to enhance the capability of the nanofiber resonator, we have demonstrated a storage solution to prevent degradation of the high-finesse resonator. The storage container effectively maintains the quality of the resonator, with a transmission loss addition under 0.08\% over a week. 
The storage container for the nanofiber resonator effectively isolates the stages of resonator fabrication and subsequent experimental studies, enabling individual optimization of each stage and broadening the range of applications.

\section{Nanofiber Fabry-P\'{e}rot resonator}

\subsection{Fiber Bragg grating mirror fabrication}
We have developed a fabrication method of FBG mirrors and fiber resonators using deep ultraviolet (DUV) laser exposure with a silica phase mask\cite{Hill1993-kc}. The specific details of this technique are discussed elsewhere\cite{Kato2022-ed}, so here we provide a brief description of the experimental setup.

We employ a 213\,nm laser source with a typical output power of 100\,mW. The DUV beam is focused on the fiber using a cylindrical lens through an optical phase mask. The diffracted beams at the phase mask create an interference pattern on the fiber core. The pitch of the interference pattern is determined by the required characteristics of the mirror. In this work, the center of the stop band of the mirror is located at 852\,nm.
The DUV beam spot can be scanned using a computer-controlled stage, allowing us to fabricate two mirrors sequentially with high position precision on the fiber. We use a standard single mode fiber (SM800(5.6/125), Fibercore) as the base fiber. The distance between the FBGs is approximately 14\,mm, with each FBG length around 10\,mm.
The typical finesse of the fabricated resonator reaches 9.0$\times10^3$ at the operating wavelength, as described in the following section.

\subsection{Fiber pulling}
To maximize the efficiency of the optical nanofiber as a waveguide, it is important to design and fabricate a properly tapered region that connects it to a standard single-mode optical fiber. We optimized the tapered geometry to balance adiabaticity of the transmission and the total length of the tapered region.
The taper shape is specifically designed for transmitting 852\,nm laser light\cite{Ruddell2020-vb}. The total length of the tapered region, including a 1\,mm nanofiber area, is approximately 27\,mm. 
The choice of nanofiber diameter depends on the application, and a diameter of 550\,nm has been utilized throughout this paper.

We manufacture the optical nanofiber using a heat-and-pull method, utilizing an oxygen-hydrogen flame as the heat source. The pulling process is controlled by computer-controlled translation stages.
The heating condition is carefully optimized to reduce the additional loss during the nanofiber fabrication.
Throughout the process, we monitor the transmittance using an intensity-stabilized laser source to characterize the pulling sequence.
The fabrication process takes place in a clean environment (class 10).

\section{Loss evaluation}
To improve the fabrication process of a high-finesse nanofiber resonator, it is crucial to accurately measure the additional loss inside the resonator at each fabrication step.
In order to evaluate the loss, we performed reflection spectroscopy from both ends of the resonator, allowing us to extract the loss and transmittance of each mirror. The reflection spectroscopy is straightforward and reliable for input power calibration, as compared to transmission spectroscopy.

The total photon loss during one round-trip in the resonator ($\alpha_{\rm all}$) can be divided into three components: $\alpha_{\rm all} = T_{\rm in} + T_{\rm end} + \alpha_{\rm loss}$. Here, $T_{\rm in (end)}$, and $\alpha_{\rm loss}$ represent the transmittance of the input (end) mirror and the round-trip transmission loss in the resonator, respectively (Fig.1{\bf a}).
The reflection spectrum provides information about $F_{\rm obs}$, $\alpha_{\rm all}$ and the ratio of $T_{\rm in}$ to $\alpha_{\rm all}$:
\begin{eqnarray}
    \frac{\rm FSR}{\rm FWHM}& = F_{\rm obs}=\frac{2\pi}{\alpha_{\rm all}}, \\
    R&=\left(1-\frac{2T_{\rm in}}{\alpha_{\rm all}}\right)^2.
\end{eqnarray}
In these equations, FSR, FWHM, $F_{\rm obs}$, and $R$ denote the free-spectral range of the resonator, the full-width half-maximum of the resonance peak, the observed finesse and the input mirror reflectance at the resonance condition, respectively. Here, we assume that the reflectance is close to unity. It is important to note that the change in FWHMs across the critical coupling regime ($R=0$) enables us to determine the over-coupled regime ($T_{\rm in} > T_{\rm end}+\alpha_{\rm loss}$) or the under-coupled regime ($T_{\rm in} < T_{\rm end}+\alpha_{\rm loss}$).
Using these relationships, the reflection spectroscopy from both the FBG mirrors allows for the calculation of the transmission loss inside the resonator ($\alpha_{\rm loss}$).

\begin{figure}
\includegraphics{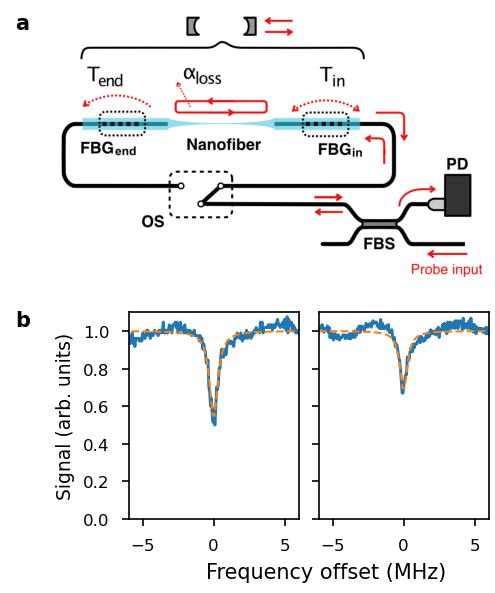}
\caption{
{\bf a} Schematic diagram of the reflection spectroscopy setup.
OS, optical switch; FBS, fiber beam splitter; PD, photodetector.
{\bf b} Example reflection spectra taken from both ports. The observed resonances are in the under-coupled regime. The orange dashed lines are Lorentzian fits to the spectra. The weak sinusoidal modulation observed in the background signal is likely attributed to an etalon effect arising from the optical fiber facets within the system.
}
\end{figure}

Figure 1{\bf a} presents a schematic setup for reflection spectroscopy. We utilize an optical switch to select an input port of the resonator and scan the probe laser frequency to record the reflection spectrum and measure the FSR.
The ${\rm FSR}$ of the nanofiber resonator is 2.35\,{\rm GHz} throughout this work.
In Fig.\,1{\bf b}, we show the reflection spectra measured from both the FBG mirrors.
We use a Lorentzian function to numerically fit the observed spectrum.
Based on the fit results, $F_{\rm obs}$ and $\alpha_{\rm loss}$ are calculated to be 3.6(2)$\times 10^3$, and 0.14(1)\%, respectively.
In addition, we introduce an intrinsic finesse $F_{\rm int}=2\pi/\alpha_{\rm loss}$, yielding 4.6(4) $\times 10^3$, in this case.
As discussed later, the mirror reflectance of the FBG depends on the operating wavelength, and it can be controlled by the FBG temperature\cite{Kato2015-ax} or applied tension.
Therefore, the $F_{\rm int}$ is a useful parameter to characterize the quality of the nanofiber resonator.
Note that the agreement of $\alpha_{\rm all}$ was confirmed through cavity ring-down measurements, where the laser linewidth does not affect measurement accuracy.

Next, we prepared another sample of a resonator and measured ${\alpha_{\rm all}}$ and ${\alpha_{\rm loss}}$ for each resonance inside the stop-bands of FBGs.
The measurement results of ${\alpha_{\rm all}}$ and ${\alpha_{\rm loss}}$ before (after) the nanofiber fabrication are summarized in the left (right) panel of Fig.\,2.
Since the large dispersion of the fabricated FBGs, the $\alpha_{\rm all}$ shows dependence on the resonance frequency due to the reflectance change along the wavelength.
On the other hand, the $\alpha_{\rm loss}$ shows less dependence.
The $\alpha_{\rm loss}$ in the resonator before the nanofiber fabrication is approximately $3.2\times10^{-4}$, and the observed finesse reaches 9.0$\times 10^3$. 
The $\alpha_{\rm loss}$ increases to approximately $2.5\times10^{-3}$ after the nanofiber fabrication. These results suggest that the loss in the nanofiber resonator is primarily due to the transmission loss added during the pulling process.
We introduce the average of the measured $\alpha_{\rm loss}$ values of resonances (${ \bar \alpha}_{\rm loss}$) to characterize the loss in the resonator for evaluating a storage method in the following section.

It should be noted that the loss inside the resonator includes the loss at the FBGs and the propagation loss in the standard fiber region.
The expected loss based on the fiber specification before the pulling process is 3.5$\times 10^{-5}$, and the expected finesse is 1.8$\times 10^5$.
The evaluated loss before the pulling process is much larger than the expected loss.
We believe that the loss is dominated by the loss at the FBGs and can be improved by using longer FBG lengths with lower refractive index contrast and better uniformity patterns.

\begin{figure}
\includegraphics{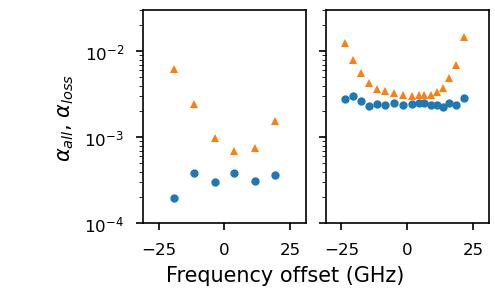}
\caption{
Evaluated losses for each resonance inside the FBG stop-bands. Orange triangles and blue circles represent $\alpha_{\rm all}$ and $\alpha_{\rm loss}$, respectively. Left (right) panel shows the results before (after) the pulling process, and the ${\rm FSR}$ is 6.85 (2.35)\,GHz, respectively.
}
\end{figure}

\section{Nanofiber resonator storage}

Optical devices that utilize evanescent fields, such as optical nanofibers, are susceptible to degradation of their optical properties due to contamination from dust or molecular deposition on their surfaces\cite{Fujiwara2011-hv, Bouhadida2021-gb}. Nanofiber devices, in particular, are subject to different experimental setups during the fabrication (fiber pulling) process and the applications such as experiments in high vacuum or cryogenic environments. Therefore, it is crucial to have a container that can transport the devices to the experimental setup without any degradation, allowing the nanofiber to function as a high-quality optical device.

In this study, we have constructed a small, portable container to maintain the nanofiber resonator under a clean nitrogen gas environment. The storage container is schematically depicted in Fig.\,3{\bf a}. Initially, the nanofiber resonator is fabricated in a clean environment and fixed onto a U-shaped jig to hold it in place. Subsequently, the jig is inserted into the container under a flow of dry, clean nitrogen gas. The optical fibers are taken out of the container through a fiber feed-through. Once the installation is complete, we close the valves in the nitrogen gas supply line and move the container from the clean environment to a standard laboratory environment. Loss evaluation measurements are conducted approximately once per day after the resonator is inserted into the container.

In Fig.\,3{\bf b}, we show the loss measurement result over the 11-day observation period.
A relatively large increase of approximately 0.08\% in ${\bar \alpha}_{\rm loss}$ was observed the day after the insertion.
We suspect contamination by small particles inside the container during the insertion process, and we are currently investigating this matter.
After the initial change, we observed stable results of ${\bar \alpha}_{\rm loss}$, and the change over the last 10 days is less than 0.03\%, and the $F_{\rm int}$ remains larger than 2.5$\times 10^3$.

\begin{figure}
\includegraphics{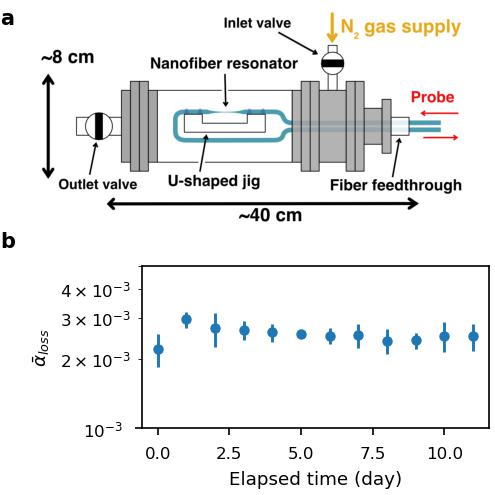}
\caption{
{\bf a} Schematic drawing of the storage container.
The container has two valves to control the nitrogen gas flow.
We performed the reflection spectroscopy via the optical fibers taken from the fiber feed-through. 
{\bf b} ${\bar \alpha}_{\rm loss}$ change in the container.
The error bar shows the standard errors of the mean. We use more than ten resonances to calculate the mean.
}
\end{figure}

\section{Conclusion}

We have developed characterization and storage methods for a high-finesse nanofiber resonator.
Using reflection spectroscopy from both ends, we evaluated $\alpha_{\rm loss}$.
The $\alpha_{\rm loss}$ for each resonance within the stop-bands of FBGs shows minimal dependence on the resonance wavelength, whereas $\alpha_{\rm all}$ demonstrates a clear dependence, indicative of the changes in mirror reflectance along the wavelength.
Furthermore, we introduced $\bar{\alpha}_{\rm loss}$ to characterize the changes in the nanofiber resonator post-fabrication.
The storage container, by maintaining the nanofiber resonator in a nitrogen gas environment, restricted the increase in loss to less than 0.03\% over 10 days, aside from the initial change of 0.08\%, and $F_{\rm int}$ remained greater than 2.5$\times 10^3$ after storage.
The observed $F_{\rm int}$ suggests a projected cooperativity parameter of approximately 20 for a CQED system when a cesium atom is trapped 300\,nm from the nanofiber surface\cite{Ruddell2020-vb}.

The demonstrated storage method holds significant practical importance for experiments involving high-finesse nanofiber resonators.
Nanofiber devices require a clean environment and careful handling during fabrication and application.
With the established storage method, the portability of the container allows for separation between the fabrication and application locations.

The flexibility provided by this storage method makes it particularly appealing for applications in quantum optics, quantum communications, and quantum sensing using nanofiber resonators.

\begin{acknowledgments}
We would like to acknowledge H. Iida, N. Shiraishi, and T. Watanabe for their help in nanofiber fabrication. We are also grateful to the NanoQT team for their kind assistance.
This work was supported by JST Moonshot R\&D (Grant Number JPMJMS2268).
\end{acknowledgments}

\hspace{1em}

{\bf Data availability}

The data that support the findings of this study are available from the corresponding author upon reasonable request.

\bibliographystyle{apsrev4-1}
\bibliography{fd_01}

\begin{thebibliography}{13}%
\makeatletter
\providecommand \@ifxundefined [1]{%
 \@ifx{#1\undefined}
}%
\providecommand \@ifnum [1]{%
 \ifnum #1\expandafter \@firstoftwo
 \else \expandafter \@secondoftwo
 \fi
}%
\providecommand \@ifx [1]{%
 \ifx #1\expandafter \@firstoftwo
 \else \expandafter \@secondoftwo
 \fi
}%
\providecommand \natexlab [1]{#1}%
\providecommand \enquote  [1]{``#1''}%
\providecommand \bibnamefont  [1]{#1}%
\providecommand \bibfnamefont [1]{#1}%
\providecommand \citenamefont [1]{#1}%
\providecommand \href@noop [0]{\@secondoftwo}%
\providecommand \href [0]{\begingroup \@sanitize@url \@href}%
\providecommand \@href[1]{\@@startlink{#1}\@@href}%
\providecommand \@@href[1]{\endgroup#1\@@endlink}%
\providecommand \@sanitize@url [0]{\catcode `\\12\catcode `\$12\catcode `\&12\catcode `\#12\catcode `\^12\catcode `\_12\catcode `\%12\relax}%
\providecommand \@@startlink[1]{}%
\providecommand \@@endlink[0]{}%
\providecommand \url  [0]{\begingroup\@sanitize@url \@url }%
\providecommand \@url [1]{\endgroup\@href {#1}{\urlprefix }}%
\providecommand \urlprefix  [0]{URL }%
\providecommand \Eprint [0]{\href }%
\providecommand \doibase [0]{http://dx.doi.org/}%
\providecommand \selectlanguage [0]{\@gobble}%
\providecommand \bibinfo  [0]{\@secondoftwo}%
\providecommand \bibfield  [0]{\@secondoftwo}%
\providecommand \translation [1]{[#1]}%
\providecommand \BibitemOpen [0]{}%
\providecommand \bibitemStop [0]{}%
\providecommand \bibitemNoStop [0]{.\EOS\space}%
\providecommand \EOS [0]{\spacefactor3000\relax}%
\providecommand \BibitemShut  [1]{\csname bibitem#1\endcsname}%
\let\auto@bib@innerbib\@empty
\bibitem [{\citenamefont {Brambilla}(2010)}]{Brambilla2010-nj}%
  \BibitemOpen
  \bibfield  {author} {\bibinfo {author} {\bibfnamefont {G.}~\bibnamefont {Brambilla}},\ }\href@noop {} {\bibfield  {journal} {\bibinfo  {journal} {J. Opt.}\ }\textbf {\bibinfo {volume} {12}},\ \bibinfo {pages} {043001} (\bibinfo {year} {2010})}\BibitemShut {NoStop}%
\bibitem [{\citenamefont {Morrissey}\ \emph {et~al.}(2013)\citenamefont {Morrissey}, \citenamefont {Deasy}, \citenamefont {Frawley}, \citenamefont {Kumar}, \citenamefont {Prel}, \citenamefont {Russell}, \citenamefont {Truong},\ and\ \citenamefont {Chormaic}}]{Morrissey2013-ra}%
  \BibitemOpen
  \bibfield  {author} {\bibinfo {author} {\bibfnamefont {M.~J.}\ \bibnamefont {Morrissey}}, \bibinfo {author} {\bibfnamefont {K.}~\bibnamefont {Deasy}}, \bibinfo {author} {\bibfnamefont {M.}~\bibnamefont {Frawley}}, \bibinfo {author} {\bibfnamefont {R.}~\bibnamefont {Kumar}}, \bibinfo {author} {\bibfnamefont {E.}~\bibnamefont {Prel}}, \bibinfo {author} {\bibfnamefont {L.}~\bibnamefont {Russell}}, \bibinfo {author} {\bibfnamefont {V.~G.}\ \bibnamefont {Truong}}, \ and\ \bibinfo {author} {\bibfnamefont {S.~N.}\ \bibnamefont {Chormaic}},\ }\href@noop {} {\bibfield  {journal} {\bibinfo  {journal} {Sensors}\ }\textbf {\bibinfo {volume} {13}},\ \bibinfo {pages} {10449} (\bibinfo {year} {2013})}\BibitemShut {NoStop}%
\bibitem [{\citenamefont {Ward}\ \emph {et~al.}(2014)\citenamefont {Ward}, \citenamefont {Maimaiti}, \citenamefont {Le},\ and\ \citenamefont {Chormaic}}]{Ward2014-bq}%
  \BibitemOpen
  \bibfield  {author} {\bibinfo {author} {\bibfnamefont {J.~M.}\ \bibnamefont {Ward}}, \bibinfo {author} {\bibfnamefont {A.}~\bibnamefont {Maimaiti}}, \bibinfo {author} {\bibfnamefont {V.~H.}\ \bibnamefont {Le}}, \ and\ \bibinfo {author} {\bibfnamefont {S.~N.}\ \bibnamefont {Chormaic}},\ }\href@noop {} {\bibfield  {journal} {\bibinfo  {journal} {Rev. Sci. Instrum.}\ }\textbf {\bibinfo {volume} {85}} (\bibinfo {year} {2014})}\BibitemShut {NoStop}%
\bibitem [{\citenamefont {Ismaeel}\ \emph {et~al.}(2013)\citenamefont {Ismaeel}, \citenamefont {Lee}, \citenamefont {Ding}, \citenamefont {Belal},\ and\ \citenamefont {Brambilla}}]{Ismaeel2013-od}%
  \BibitemOpen
  \bibfield  {author} {\bibinfo {author} {\bibfnamefont {R.}~\bibnamefont {Ismaeel}}, \bibinfo {author} {\bibfnamefont {T.}~\bibnamefont {Lee}}, \bibinfo {author} {\bibfnamefont {M.}~\bibnamefont {Ding}}, \bibinfo {author} {\bibfnamefont {M.}~\bibnamefont {Belal}}, \ and\ \bibinfo {author} {\bibfnamefont {G.}~\bibnamefont {Brambilla}},\ }\href@noop {} {\bibfield  {journal} {\bibinfo  {journal} {Laser Photon. Rev.}\ }\textbf {\bibinfo {volume} {7}},\ \bibinfo {pages} {350} (\bibinfo {year} {2013})}\BibitemShut {NoStop}%
\bibitem [{\citenamefont {Tong}(2018)}]{Tong2018-fh}%
  \BibitemOpen
  \bibfield  {author} {\bibinfo {author} {\bibfnamefont {L.}~\bibnamefont {Tong}},\ }\href@noop {} {\bibfield  {journal} {\bibinfo  {journal} {Sensors}\ }\textbf {\bibinfo {volume} {18}} (\bibinfo {year} {2018})}\BibitemShut {NoStop}%
\bibitem [{\citenamefont {Praveen~Kamath}\ \emph {et~al.}(2023)\citenamefont {Praveen~Kamath}, \citenamefont {Sil}, \citenamefont {Truong},\ and\ \citenamefont {Nic~Chormaic}}]{Praveen_Kamath2023-wl}%
  \BibitemOpen
  \bibfield  {author} {\bibinfo {author} {\bibfnamefont {P.}~\bibnamefont {Praveen~Kamath}}, \bibinfo {author} {\bibfnamefont {S.}~\bibnamefont {Sil}}, \bibinfo {author} {\bibfnamefont {V.~G.}\ \bibnamefont {Truong}}, \ and\ \bibinfo {author} {\bibfnamefont {S.}~\bibnamefont {Nic~Chormaic}},\ }\href@noop {} {\bibfield  {journal} {\bibinfo  {journal} {Biomed. Opt. Express}\ }\textbf {\bibinfo {volume} {14}},\ \bibinfo {pages} {6172} (\bibinfo {year} {2023})}\BibitemShut {NoStop}%
\bibitem [{\citenamefont {Kato}\ and\ \citenamefont {Aoki}(2015)}]{Kato2015-ax}%
  \BibitemOpen
  \bibfield  {author} {\bibinfo {author} {\bibfnamefont {S.}~\bibnamefont {Kato}}\ and\ \bibinfo {author} {\bibfnamefont {T.}~\bibnamefont {Aoki}},\ }\href@noop {} {\bibfield  {journal} {\bibinfo  {journal} {Phys. Rev. Lett.}\ }\textbf {\bibinfo {volume} {115}},\ \bibinfo {pages} {093603} (\bibinfo {year} {2015})}\BibitemShut {NoStop}%
\bibitem [{\citenamefont {Ruddell}\ \emph {et~al.}(2017)\citenamefont {Ruddell}, \citenamefont {Webb}, \citenamefont {Herrera}, \citenamefont {Parkins},\ and\ \citenamefont {Hoogerland}}]{Ruddell2017-uo}%
  \BibitemOpen
  \bibfield  {author} {\bibinfo {author} {\bibfnamefont {S.~K.}\ \bibnamefont {Ruddell}}, \bibinfo {author} {\bibfnamefont {K.~E.}\ \bibnamefont {Webb}}, \bibinfo {author} {\bibfnamefont {I.}~\bibnamefont {Herrera}}, \bibinfo {author} {\bibfnamefont {A.~S.}\ \bibnamefont {Parkins}}, \ and\ \bibinfo {author} {\bibfnamefont {M.~D.}\ \bibnamefont {Hoogerland}},\ }\href@noop {} {\bibfield  {journal} {\bibinfo  {journal} {Optica}\ }\textbf {\bibinfo {volume} {4}},\ \bibinfo {pages} {576} (\bibinfo {year} {2017})}\BibitemShut {NoStop}%
\bibitem [{\citenamefont {Fujiwara}\ \emph {et~al.}(2011)\citenamefont {Fujiwara}, \citenamefont {Toubaru}, \citenamefont {Noda}, \citenamefont {Zhao},\ and\ \citenamefont {Takeuchi}}]{Fujiwara2011-hv}%
  \BibitemOpen
  \bibfield  {author} {\bibinfo {author} {\bibfnamefont {M.}~\bibnamefont {Fujiwara}}, \bibinfo {author} {\bibfnamefont {K.}~\bibnamefont {Toubaru}}, \bibinfo {author} {\bibfnamefont {T.}~\bibnamefont {Noda}}, \bibinfo {author} {\bibfnamefont {H.-Q.}\ \bibnamefont {Zhao}}, \ and\ \bibinfo {author} {\bibfnamefont {S.}~\bibnamefont {Takeuchi}},\ }\href@noop {} {\bibfield  {journal} {\bibinfo  {journal} {Nano Lett.}\ }\textbf {\bibinfo {volume} {11}},\ \bibinfo {pages} {4362} (\bibinfo {year} {2011})}\BibitemShut {NoStop}%
\bibitem [{\citenamefont {Bouhadida}\ \emph {et~al.}(2021)\citenamefont {Bouhadida}, \citenamefont {Delaye},\ and\ \citenamefont {Lebrun}}]{Bouhadida2021-gb}%
  \BibitemOpen
  \bibfield  {author} {\bibinfo {author} {\bibfnamefont {M.}~\bibnamefont {Bouhadida}}, \bibinfo {author} {\bibfnamefont {P.}~\bibnamefont {Delaye}}, \ and\ \bibinfo {author} {\bibfnamefont {S.}~\bibnamefont {Lebrun}},\ }\href@noop {} {\bibfield  {journal} {\bibinfo  {journal} {Opt. Commun.}\ }\textbf {\bibinfo {volume} {500}},\ \bibinfo {pages} {127336} (\bibinfo {year} {2021})}\BibitemShut {NoStop}%
\bibitem [{\citenamefont {Ruddell}\ \emph {et~al.}(2020)\citenamefont {Ruddell}, \citenamefont {Webb}, \citenamefont {Takahata}, \citenamefont {Kato},\ and\ \citenamefont {Aoki}}]{Ruddell2020-vb}%
  \BibitemOpen
  \bibfield  {author} {\bibinfo {author} {\bibfnamefont {S.~K.}\ \bibnamefont {Ruddell}}, \bibinfo {author} {\bibfnamefont {K.~E.}\ \bibnamefont {Webb}}, \bibinfo {author} {\bibfnamefont {M.}~\bibnamefont {Takahata}}, \bibinfo {author} {\bibfnamefont {S.}~\bibnamefont {Kato}}, \ and\ \bibinfo {author} {\bibfnamefont {T.}~\bibnamefont {Aoki}},\ }\href@noop {} {\bibfield  {journal} {\bibinfo  {journal} {Opt. Lett.}\ }\textbf {\bibinfo {volume} {45}},\ \bibinfo {pages} {4875} (\bibinfo {year} {2020})}\BibitemShut {NoStop}%
\bibitem [{\citenamefont {Hill}\ \emph {et~al.}(1993)\citenamefont {Hill}, \citenamefont {Malo}, \citenamefont {Bilodeau}, \citenamefont {Johnson},\ and\ \citenamefont {Albert}}]{Hill1993-kc}%
  \BibitemOpen
  \bibfield  {author} {\bibinfo {author} {\bibfnamefont {K.~O.}\ \bibnamefont {Hill}}, \bibinfo {author} {\bibfnamefont {B.}~\bibnamefont {Malo}}, \bibinfo {author} {\bibfnamefont {F.}~\bibnamefont {Bilodeau}}, \bibinfo {author} {\bibfnamefont {D.~C.}\ \bibnamefont {Johnson}}, \ and\ \bibinfo {author} {\bibfnamefont {J.}~\bibnamefont {Albert}},\ }\href@noop {} {\bibfield  {journal} {\bibinfo  {journal} {Appl. Phys. Lett.}\ }\textbf {\bibinfo {volume} {62}},\ \bibinfo {pages} {1035} (\bibinfo {year} {1993})}\BibitemShut {NoStop}%
\bibitem [{\citenamefont {Kato}\ and\ \citenamefont {Aoki}(2022)}]{Kato2022-ed}%
  \BibitemOpen
  \bibfield  {author} {\bibinfo {author} {\bibfnamefont {S.}~\bibnamefont {Kato}}\ and\ \bibinfo {author} {\bibfnamefont {T.}~\bibnamefont {Aoki}},\ }\href@noop {} {\bibfield  {journal} {\bibinfo  {journal} {Opt. Lett.}\ }\textbf {\bibinfo {volume} {47}},\ \bibinfo {pages} {5000} (\bibinfo {year} {2022})}\BibitemShut {NoStop}%
\end{thebibliography}%

\end{document}